\documentclass[a4paper,fleqn,usenatbib]{mnras}

\usepackage{newtxtext,newtxmath}

\usepackage[T1]{fontenc}
\usepackage{ae,aecompl}

\usepackage{graphicx}	
\usepackage{amsmath}	
\usepackage{amssymb}	

\title[Two stellar/substellar binaries in Praesepe]{Two close binaries across the hydrogen-burning limit in the Praesepe open cluster}

\author[N. Lodieu et al.]{N.\ Lodieu$^{1,2}$\thanks{E-mail: nlodieu@iac.es (NL)},
C.\ del Burgo$^{3,1}$,
E.\ Manjavacas$^{4,5}$,
M.\ R.\ Zapatero Osorio$^{6}$
C. Alvarez$^{4}$,
\newauthor
V. J. S. B\'ejar$^{1,2}$,
S.\ Boudreault$^{7}$,
J.\ Lyke$^{4}$,
R. Rebolo$^{1,2,8}$,
P. Chinchilla$^{1,2}$
\newauthor
\\
$^{1}$Instituto de Astrof\'isica de Canarias Calle V\'ia L\'actea S/N 38205 La Laguna, Tenerife, Spain\\
$^{2}$Departamento de Astrof\'isica, Universidad de La Laguna (ULL), E-38206 La Laguna, Tenerife, Spain\\
$^{3}$Instituto Nacional de Astrof\'isica, \'Optica y Electr\'onica, Luis Enrique Erro 1, Sta.\ Ma.\ Tonantzintla, Puebla, Mexico\\
$^{4}$W. M. Keck Observatory, 65-1120 Mamalahoa Hwy, Kamuela, HI, 96743, USA\\
$^{5}$Department of Astronomy/Steward Observatory, The University of Arizona, 933 N. Cherry Avenue, Tucson, AZ 85721, USA\\
$^{6}$Centro de Astrobiolog\'ia (CSIC/INTA), 28850 Torrej\'on de Ardoz, Madrid, Spain\\
$^{7}$Max-Planck-Institut f\"ur Sonnensystemforschung, Justus-von-Liebig-Weg 3, 37077, G\"ottingen, Germany\\
$^{8}$Consejo Superior de Investigaciones Cient\'ificas, CSIC, Spain\\
}

\date{Accepted XXX. Received YYY; \today{}}

\pubyear{2019}

\begin{document}
\label{firstpage}
\pagerange{\pageref{firstpage}--\pageref{lastpage}}
\maketitle

\begin{abstract}
We present Keck\,I/OSIRIS and Keck\,II/NIRC2 adaptive optics imaging of two member candidates
of the Praesepe stellar cluster (d\,=\,186.18$\pm$0.11 pc; 590--790 Myr), UGCS\,J08451066+2148171
(L1.5$\pm$0.5) and UGCS\,J08301935$+$2003293 (no spectroscopic classification). We resolved UGCS\,J08451066$+$2148171 
into a binary system in the near-infrared, with a $K$-band wavelength flux ratio of 0.89$\pm$0.04,
a projected separation of 60.3$\pm$1.3 mas (11.2$\pm$0.7 au; 1$\sigma$).
We also resolved UGCS\,J08301935$+$2003293 into a binary system with a flux ratio of 
0.46$\pm$0.03 and a separation of 62.5$\pm$0.9 mas.
Assuming zero eccentricity, we estimate minimum orbital periods of $\sim$100 years for both systems.
According to theoretical evolutionary models, we derive masses in the range of 
0.074--0.078 M$_{\odot}$ and
0.072--0.076 M$_{\odot}$ for the primary and secondary of UGCS\,J08451066$+$2148171
for an age of 700$\pm$100 Myr. In the case of 
UGCS\,J08301935$+$2003293, the primary is a low-mass star at the stellar/substellar boundary 
(0.070--0.078 M$_{\odot}$) while the companion candidate might be a brown dwarf (0.051--0.065 M$_{\odot}$).
These are the first two binaries composed of L dwarfs in Praesepe. They are benchmark systems to
derive the location of the substellar limit at the age and metallicity of Praesepe, determine the age 
of the cluster based on the lithium depletion boundary test, derive dynamical masses, and improve
low-mass stellar and substellar evolutionary models at a well-known age and metallicity.
\end{abstract}

\begin{keywords}
(stars:) binaries: general -- stars: late-type -- open clusters: Praesepe
\end{keywords}

%
%
\section{Introduction}
\label{Prae_L0_bin:intro}

The fate of a star is primarily set by its mass. 
Dynamical masses of single and multiple stellar and substellar systems serve as valuable 
input to constrain theoretical evolutionary models and to compare with predictions of 
physical parameters by star and brown dwarf formation models \citep{zapatero04b,delBurgo18}.
Brown dwarfs cool down as they age, meaning that their masses are challenging to determine
without precise values for their age. Hence, studying multiple systems at the hydrogen-burning limit at different ages and
metallicities is of prime interest to derive dynamical masses, locate the stellar/substellar
boundary, and calibrate evolutionary models.

The number of low-mass stars with wide ultracool and substellar companions has significantly
increased over the past decades with independent dedicated surveys in star-forming regions
\citep{ahmic07,kraus06,todorov10a}, OB associations \citep{kraus05,bouy06b,biller11}, young moving
groups \citep{naud17a}, intermediate-age clusters such as the Pleiades \citep{martin00a,martin03,bouy06a}, 
and older clusters like the Hyades \citep{duchene13a}, and in the field \citep[e.g.][]{burgasser07a,dupuy17}. 
Additional discoveries have been reported as part of studies involving smaller sample sizes in 
diverse regions such as Orion \citep{stassun06}, Taurus \citep{todorov10a,konopacky07b}, 
LkH$\alpha$233 \citep{allers09}, Chamaeleon \citep{joergens06a,joergens07,joergens08}, 
Upper Scorpius \citep{bejar08,chinchilla20a}, R Corona Australis \citep{bouy04b}, 
TW Hydra \citep{chauvin05b,mohanty07}, AB Doradus \citep{desrochers18}, and Tucana-Horologium \citep{artigau15a}.
These studies, among others, seem to point towards a gradual decline of multiplicity with mass and limited
differences as a function of age, although not all surveys map the same mass and separation intervals.
Nonetheless, binary systems are of prime importance to scale masses as a function of age and
improve stellar evolutionary models \citep[e.g.][]{baraffe15}.

The Praesepe cluster (a.k.a.\ the Beehive cluster, M44) is relatively young and one of the nearest clusters to the Sun.
It is located at a distance of 186.18$\pm$0.11 pc \citep{babusiaux18,Gaia_Brown2018,Gaia_Lindegren2018} 
with a tidal radius of 10.7 pc \citep{lodieu19b}. We adopt these
values throughout the analysis presented in this paper. The range in age,
590--790 Myr, is debated in the literature \citep{mermilliod81,fossati08,delorme11,brandt15b,gossage18}, 
but comparable to the Hyades \citep{maeder81,brandt15a,martin18a,lodieu18a}.
Members of Praesepe share a significant proper motion ($\mu_{\alpha}cos(\delta)$, $\mu_{\delta}$) allowing an astrometric selection
which coupled with photometry led to a pre-$Gaia$ census of more than 1100 member candidates from high-mass
stars down to the substellar regime
\citep{hambly95,adams02,kraus07d,pinfield97,chappelle05,gonzales_garcia06,boudreault10,baker10,wang11,boudreault12,khalaj13,wang14a}.

\begin{figure*}
  \centering
  \includegraphics[width=0.48\linewidth, angle=0]{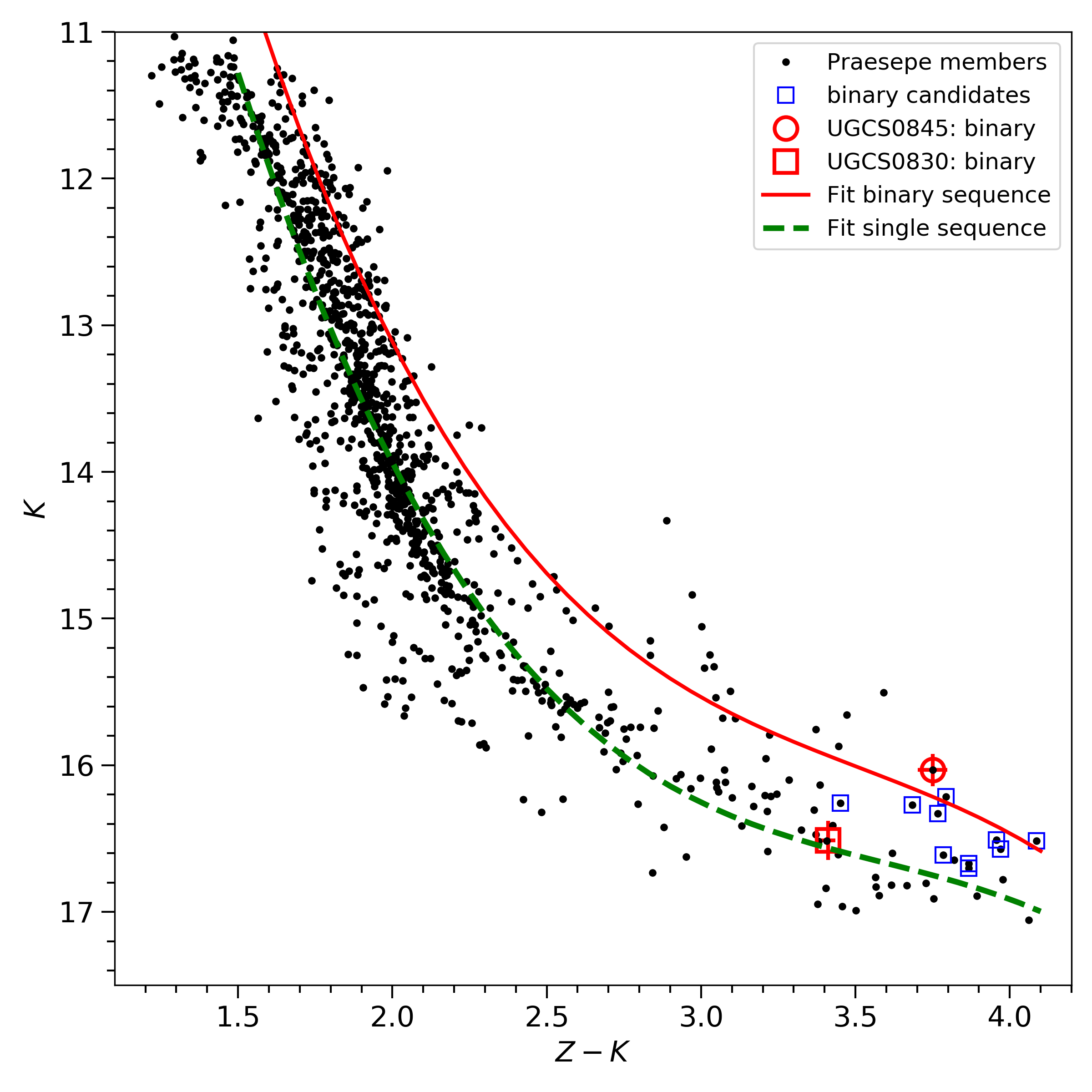}
  \includegraphics[width=0.48\linewidth, angle=0]{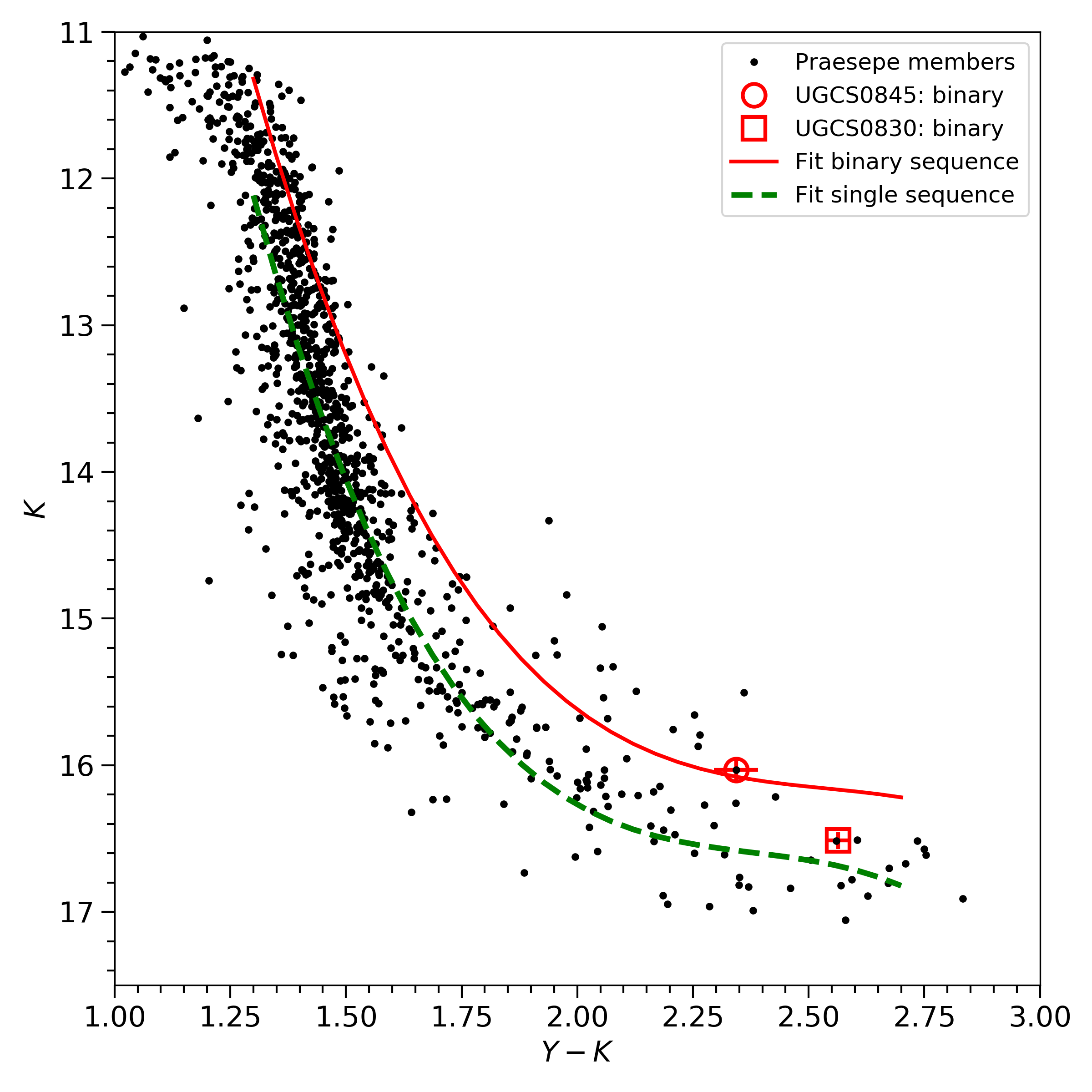}
   \caption{
{\it{Left:}} ($Z-K$,$K$) colour-magnitude diagram showing the position of UGCS0845 (red circle)
and UGCS0830 (red square) close to the binary sequence (red line) of the cluster, which contains 
over 1100 members \citep[black dots;][]{boudreault12}.
{\it{Right:}} ($Y-K$,$K$) colour-magnitude diagram.
The green and red lines represent fits to the single and equal-mass binary sequences of the cluster.
   \label{L0_Prae_binary:fig_CMDs}
   }
\end{figure*}

\citet{boudreault12} presented a pre-$Gaia$ updated census of photometric and astrometric cluster 
member candidates down to the hydrogen-burning limit in 36 square degrees of Praesepe surveyed 
by the UKIRT Infrared Deep Sky Survey Galactic Clusters Survey \citep{lawrence07}.
In a sample of over 1100 member candidates, \citet*{boudreault13} confirmed spectroscopically
the first L-type member, UGCS\,J084510.66+214817.1 (hereafter UGCS0845),
classified as an L0.3$\pm$0.4 dwarf with an effective temperature of $\sim$2300\,K and
a mass of 71.1$\pm$23.0 M$_{\rm Jup}$. Moreover, they proposed UGCS0845 as a 
photometric binary candidate because of its location above the cluster sequence in 
several colour-magnitude diagrams (Fig.\ \ref{L0_Prae_binary:fig_CMDs}).
Other member candidates lie in the potential binary sequence of Praesepe.

In this paper, we present high spatial resolution imaging of UGCS0845 and UGCS\,J08301935$+$2003293 
(hereafter UGCS0830). They have been confirmed as members of the Praesepe cluster based on their 
astrometry and photometry, and the latter also spectroscopically. In Section \ref{Prae_L0_bin:spectro}
we present optical and near-infrared spectroscopy of UGCS0845\@.
In section \ref{Prae_L0_bin:AO_LGS} we describe the adaptive optics campaign confirming both 
sources as visual binaries. In Section \ref{Prae_L0_bin:param} we infer their physical parameters, 
including separations, bolometric luminosities, and masses.
We put our results in context and discuss their impact in Section \ref{Prae_L0_bin:discussion}.

%
%
%

%
%
\begin{figure*}
  \centering
  \includegraphics[width=0.48\linewidth, angle=0]{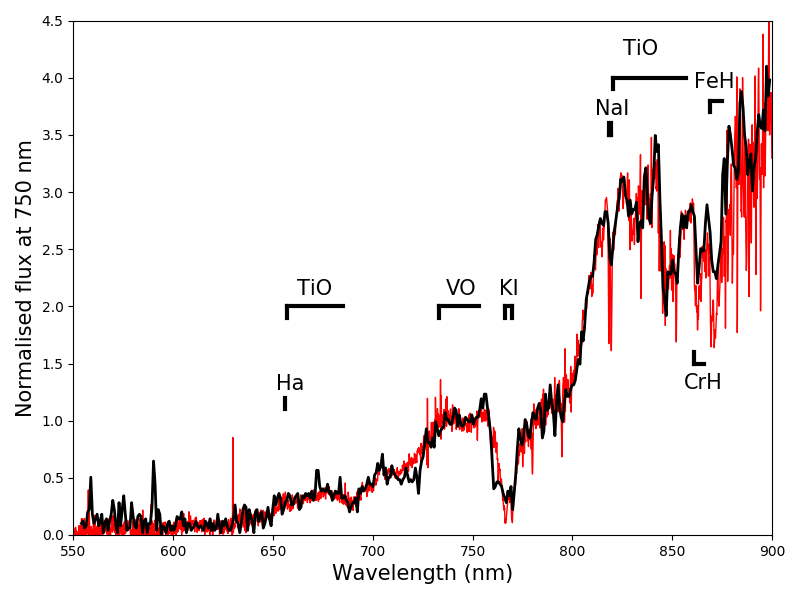}
  \includegraphics[width=0.48\linewidth, angle=0]{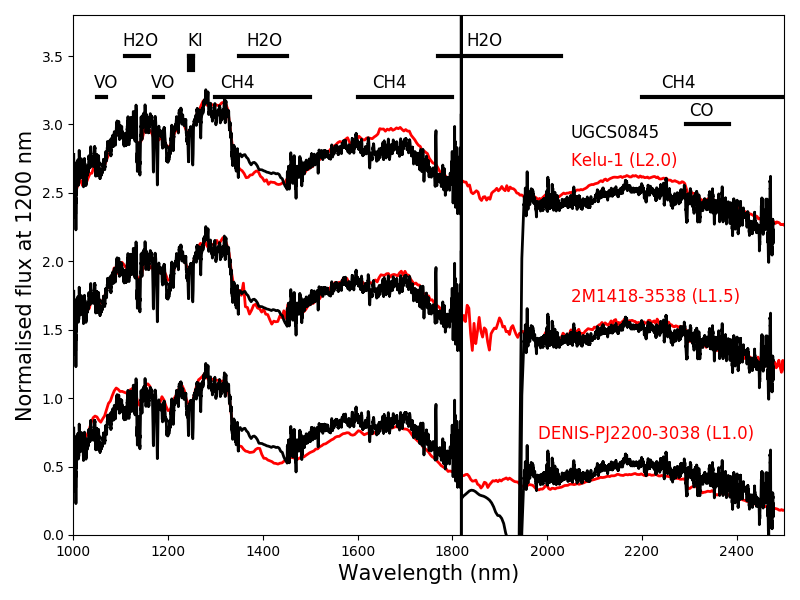}
   \caption{
{\it{Left:}} GTC/OSIRIS low-resolution optical spectrum of UGCS0845 (black) compared with the
best fit (L1.0; red) from the Sloan database of L dwarfs \citep{schmidt10b}.
The GTC spectrum is not corrected for telluric absorption while the Sloan one is.
{\it{Right:}} VLT/X-shooter near-infrared spectrum of UGCS0845 (black) compared to three L
dwarf very low resolution spectra \citep{ruiz97,kendall04,burgasser06e,burgasser07d,kirkpatrick10} 
downloaded from the SpeX archive (red). Main spectral features are highlighted.
   \label{L0_Prae_binary:fig_spectra}
   }
\end{figure*}
%

%
%
\section{Optical and near-infrared spectroscopy}
\label{Prae_L0_bin:spectro}
\subsection{Near-infrared spectroscopy}
\label{Prae_L0_bin:spectro_XSH}

We observed UGCS0845 with the X-shooter spectrograph \citep{dOdorico06,vernet11} on the European Southern 
Observatory (ESO) Very Large Telescope arrays (VLT) on the night of 01 January 2017 between 6h and 7h UT as part 
of programme number 098.C-0277(A) (P.I.\ Manjavacas). The sky was clear with a seeing of 0.6--0.9 arcsec
and the moon illuminated at 65\%.

X-shooter is composed of three arms covering the ultraviolet (UVB; 300--550 nm), visible (VIS; 550--1000 nm),
and near-infrared (NIR; 1000--2500 nm). We used slits of 1.3 arcsec in the UVB and 1.2 arcsec in the VIS and
NIR, yielding resolving powers of $\sim$2030 in the UVB, $\sim$3360 in the VIS, and $\sim$3900 in the NIR\@.
We collected 10 individual frames of 300\,s with an AB pattern and an offset of 6 arcsec to 
optimise sky subtraction in the NIR\@. We observed at parallactic angle to mitigate the effect 
of differential chromatic refraction. Due to the low signal to noise in the UVB and VIS arms, 
we only analyse the NIR spectrum. We observed the telluric standard star
HIP\,026545 at a similar airmass just after our target. Bias, flats, and arc lamps were collected
as part of the ESO calibration plan.

We reduced the spectra with the X-Shooter pipeline version 1.3.7 \citep{goldoni06,modigliani10}, which
deals with the main instrumental effects to produce a 2D combined image for each arm.
We optimally extracted the NIR spectrum with the task {\tt{apall}} in IRAF 
\citep{tody86,tody93}\footnote{IRAF is distributed by the National Optical Astronomy Observatory, which 
is operated by the Association of Universities for Research in Astronomy (AURA) under a cooperative agreement 
with the National Science Foundation}. We used the spectrum of the telluric standard to obtain the instrument
response function and correct for telluric features.
Then, we derived a response function by dividing the non-flux calibrated spectrum of the telluric standard 
(cleaned of cosmic rays and strong telluric lines) by a blackbody synthetic spectrum with the same temperature 
as the B-type telluric \citep{theodossiou91}.  Finally, we divided the spectrum of UGCS0845 by this response 
function to correct for the instrumental response and telluric lines. We display the X-shooter NIR spectrum in 
the right-hand side panel of Fig.\ \ref{L0_Prae_binary:fig_spectra}.


%
\subsection{Optical spectroscopy}
\label{Prae_L0_bin:spectro_OSIRIS}

We obtained a low-resolution optical spectrum of UGCS0845 on 16 January 2013 with the OSIRIS 
instrument \citep[Optical System for Imaging and low Resolution Integrated Spectroscopy;][]{cepa00}
mounted on the 10.4 m Gran Telescopio Canarias (GTC) in the Roque de Los Muchachos Observatory 
in La Palma (Canary Islands) under program GTC66-12B (PI Boudreault). The observing conditions
were spectroscopic (thin cirrus) with a seeing better than 1.2 arcsec and no moon.
We used the 2$\times$2 binning mode of OSIRIS with the R300R grism and a 1.2 arcsec slit width, 
yielding a resolving power of $\sim$350 at 685 nm.
This configuration shows contamination from the second-order light redwards of 850 nm,
although it becomes stronger at $>$900\,nm. We did not correct for this contamination,
hence limiting our wavelength range to 550--900 nm.

We collected six spectra with an integration time of 700\,s each, and applied an offset of 
10 arcsec along the slit to ease removal of artifacts and cosmic rays.
We reduced the spectrum in a standard manner, removing bias and flat-field from the
2D images, combining the six spectra, and optimally extracting the combined spectrum with {\tt{apall}} 
as in the case of X-shooter. We calibrated the spectrum in wavelength with a combination of HgAr, 
Ne, and Xe lamps with a rms better than 0.5\,\AA{}. We applied the response function to UGCS0845
using the spectro-photometric standard star Ross\,640 \citep{oke90} observed with the same set-up
on the same night as our target. The final spectrum, uncorrected for telluric contribution, 
is presented in the left-hand panel of Fig.\ \ref{L0_Prae_binary:fig_spectra}
along with the best fit using an Sloan L dwarf template \citep{schmidt10b}.

%
%
\section{Laser guide star adaptive optics imaging}
\label{Prae_L0_bin:AO_LGS}
\subsection{Keck/OSIRIS observations}
\label{Prae_L0_bin:AO_LGS_OSIRIS}

We observed UGCS0845 on 20 November 2018 using the OH- Suppressing Infra-Red Imaging Spectrograph
(OSIRIS) instrument \citep{larkin06a} on the Keck\,I telescope. We obtained a second epoch on 24 January 2019\@.
OSIRIS is a near-IR imager and integral field spectrograph coupled 
to the Keck Adaptive Optics (AO) system \citep{wizinowich06a,vanDam06}. Observations were performed in 
Laser-Guide-Star (LGS) mode with the recently-upgraded imaging arm of OSIRIS \citep{arriaga18}.
The OSIRIS imager has a pixel scale of $\sim$0.010 arcsec and a field of view of 20$"$\,$\times$\,20$"$. 
We used the $K_{p}$ filter ($\lambda_c = 2.144 \mu m$, $\Delta\lambda = 0.307 \mu m$) on a 
five-point dither pattern with an integration time of 57.5 seconds per position.
Additionally, we observed the astrometric binary ADS 3279 AB with the $K_{cont}$ filter
($\lambda_c = 2.270 \mu m$, $\Delta\lambda = 0.02 \mu m$) on 22 November 2018 to determine the plate 
scale of the detector with good precision. The integration time on ADS\,2279\,AB was 1.475 seconds. 
The sky was clear and the seeing below 0.8 arcsec on both nights.

Additionally, we observed UGCS0845 and UGCS0830 on 22 March 2019 with Keck\,I/OSIRIS\@. 
The sky was clear and the seeing sub-arcsec. For each target, we
collected five exposures of 15 seconds each with the $K_{p}$ filter using a small dither pattern.

%
%
\begin{figure*}
  \centering
   \includegraphics[width=0.48\linewidth, angle=0]{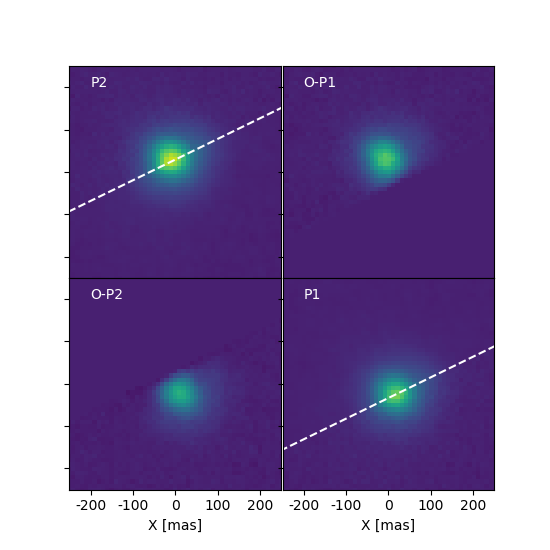}
   \includegraphics[width=0.48\linewidth, angle=0]{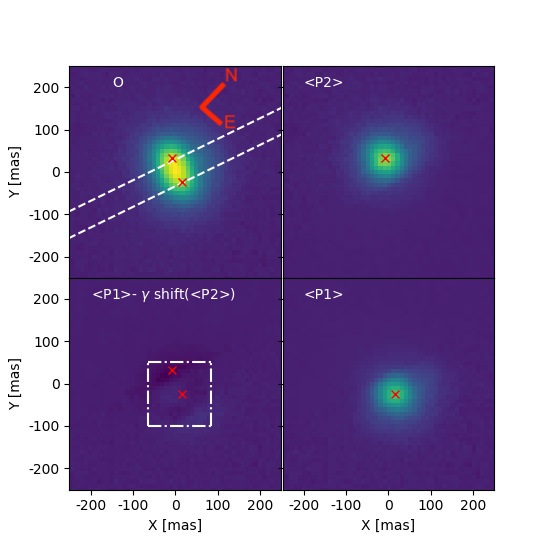}
   \caption{
{\it{Left panel:}} Illustration of the PSF decomposition method used for both instruments.
The top-left plot shows the OSIRIS image of the secondary P2, obtained from taking the signal over the dashed 
line and under the dashed line in the bottom-right plot, but with the signal shifted to the position 
and multiplied by the flux ratio. The bottom-right plot illustrates the same but for the primary.
The top-right (bottom-left) plot displays the result of subtracting image P1 (P2) to image O, 
providing a second version for the secondary (primary).
{\it{Right panel:}} Reconstruction of the UGCS0845 system.
The original OSIRIS image (O) is shown on the top-left plot. The dashed lines, parallel to each other, indicate
the limits introduced to reconstruct the PSF\@. The best reconstructed images of the secondary ($<$P2$>$)
and primary ($<$P1$>$) are displayed in the top-right and bottom-right plots with their centroid marked 
with a cross. The difference between the intensities of the secondary and the primary (red crosses), 
after being scaled by the flux ratio and shifted to the position of the secondary
is illustrated in the bottom-left plot. The box shows the region where the chi$^{2}$ statistics are calculated.
   }
   \label{L0_Prae_binary:fig_PSFimage_method}
\end{figure*}
%

%
%
\begin{figure*}
  \centering
  \includegraphics[width=0.48\linewidth, angle=0]{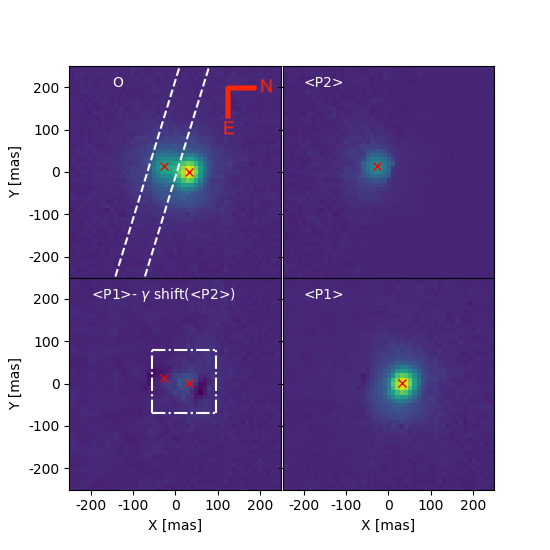}
  \includegraphics[width=0.48\linewidth, angle=0]{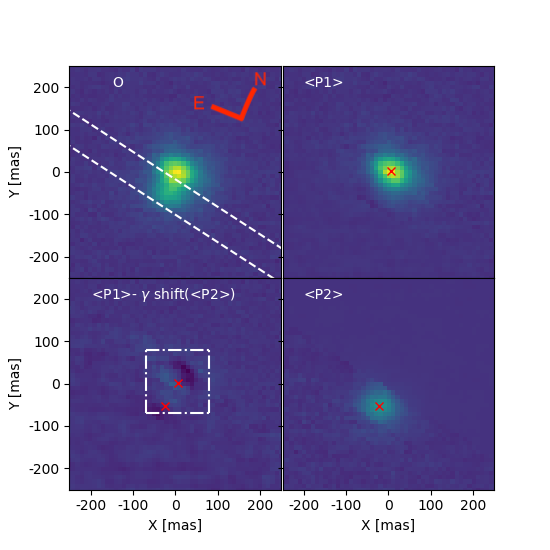}
   \caption{
{\it{Left panel:}}
{\it{Top-left:}} Reduced OSIRIS image of UGCS0830 (O). The dashed lines, parallel to each other, indicate 
the limits introduced to reconstruct the PSF of the UGCS0830 system. 
{\it{Top-right:}} Best reconstructed image of the secondary ($<$P2$>$) and its centroid marked with a red cross. 
{\it{Bottom-right:}} Best reconstructed image of the primary ($<$P1$>$) and its centroid marked with a red cross.
{\it{Bottom-left:}} Difference between the intensities of the secondary and the primary, after being 
scaled by the flux ratio and shifted to the position of the secondary.
The positions of P1 and P2 are marked with red crosses. The white box shows the region
where the chi$^{2}$ statistics are calculated.
{\it{Right panel:}} Same as left panel but for the NIRC2 images of UGCS0830\@.
We indicate the orientation of the detector with respect to North and East.
   }
   \label{L0_Prae_binary:fig_PSFimage_binaries}
\end{figure*}

We reduced the data with IRAF routines \citep{tody86,tody93} following standard procedures in the
near-infrared. We obtained uniformly illuminated images by dome lights and combined them to create a flat-field image. 
We flat-fielded the raw frames of the targets and later combined them with a median filter to remove 
the stars and create a sky background image. We sky-subtracted these images, aligned them, and combined them
to produce the final deep image.

The offsets in the dithering pattern of the images collected for UGCS0830 turned out to be too small
for a proper sky subtraction. Hence, we directly used the original images, without additional treatment.

To measure the plate scale,
we used three independent images of the binary ADS\,3297\,AB (HD\,28867) obtained on 2018 November 22 
and the astrometry given by $Gaia$ DR2, which is consistent with that of \citet{scardia07a}, although with 
significantly higher precision. According to $Gaia$, the projected separation of the two stars 
is 3.06823$\pm$0.00010 arcsec. We calculated a pixel size of the Keck/OSIRIS detector of 
9.9407$\pm$0.0071 mas, assuming squared pixels. The uncertainty accounts for the dispersion of the 
measurements.

\subsection{Keck/NIRC2 observations}
\label{Prae_L0_bin:AO_LGS_NIRC2}

We observed UGCS0845 and UGCS0830 with the NIRC2 instrument on the Keck\,II telescope in LGS-AO mode on 
11 March 2020 between UT 06h40m and 07h10m. The sky was clear and the seeing was sub-arcsec. We employed 
the $K_{s}$ filter and the NIRC2 narrow field camera with a pixel scale of 9.971$\pm$0.004 mas \citep{service16a}
and a field of view of 10$\times$10 arcsec$^{2}$. Observations were taken following a 3-point dither pattern 
with offsets of five arcsec, a jitter of a few pixels, and an on-source integration time of 60s per dither position. 
This pattern was repeated three times, thus yielding a total exposure time of nine min per target. We avoided the 
lower left quadrant of the NIRC2 1024$\times$1024 Aladdin-3 InSb detector, since it is known to be slightly 
noisier than the other three quadrants. 
We note that the images of UGCS0845 suffer from elongation due to wind shake in the direction of the 
telescope elevation at the time of the observations, making this dataset not usable.

We reduced the data in a standard manner under the IRAF environment. We median-combined
the nine raw frames per target to create a sky image. We subtracted this sky frame from each individual 
raw image before aligning and stacking all data to produce the final, deep image.

We also observed the astrometric binary ADS 7878 AB on 11 March 2020. We used the position angle of 
ADS\,7878\,AB from \citet{scardia07a}, 161.9$\pm$0.3 deg, to determine the orientation of the 
NIRC2 FOV with respect to the celestial equatorial system. 

%
%
%
\section{Physical parameters}
\label{Prae_L0_bin:param}
\subsection{Method: PSF decomposition}
\label{Prae_L0_bin:param_Method}

The separation of the two components of the systems presented in this work is below the 
full width half maximum (FWHM) of single objects observed with the same correction.
We apply a method based on point-spread function (PSF) reconstruction to estimate the most
likely flux ratio, FWHM, the angle between the line defined by the centroids of the binary components 
and a reference axis (vertical), and the angular separation of the components.
We assumed that the PSF is the same at the positions 
of the binary components (for the two cases discussed here) given their small angular separation 
on the OSIRIS detector (Fig.\ \ref{L0_Prae_binary:fig_PSFimage_method}). 

The method outlined in Fig.\ \ref{L0_Prae_binary:fig_PSFimage_method} is able to separate the
two components of the binaries. In both cases the primary (hereinafter P1) is below the secondary (P2).
The image of each component was
reconstructed after joining the flux signal above the top line with that below the bottom line, conveniently 
scaled and shifted. For example, for the primary, we leave untouched the signal below the bottom line and 
then shifted and scaled the signal above the top line to complete its shape. We applied a bi-linear 
interpolation method to shift the data (left in Fig.\ \ref{L0_Prae_binary:fig_PSFimage_method}). 

We apply the aforementioned approach taking into account four free parameters: the
flux ratio between the secondary and the primary, the multiplicative factor $\gamma$ with values
between 0 and 1, the angle of the straight lines (parallel between them) with 
respect to the x axis, and the zero points in the y axis for the upper and lower straight lines.

After reconstructing the image of a given component (for every trial), we subtracted it from the original 
image (O). In this way, we derive the image of the companion in a different way. We took the average of the 
two versions for every component. 
The method attempts to reduce the mutual flux contamination between the two sources.
Note the sum of $<$P1$>$ and $<$P2$>$ is exactly O ($\equiv$\,$<$P1$>$\,$+$\,$<$P2$>$), so we separate 
the original reduced image into two different components. We finally shift $<$P1$>$, multiply by $\gamma$,
to the position of $<$P2$>$ and then calculate the residuals. 

We explore the parameter space to find the best solution, i.e.\ the one with the smallest value of $\chi_2$
in the OSIRIS images. We also repeated the procedure on the individual sky-corrected NIRC2 images
(right panel in Fig.\ \ref{L0_Prae_binary:fig_PSFimage_binaries}).
We illustrate the method applied to separate the point-like sources of both binaries in the left-hand side
of Fig.\ \ref{L0_Prae_binary:fig_PSFimage_method}.
We show the decomposition of the best solution for UGCS0845 and UGCS0830 in the right panel of
Fig.\ \ref{L0_Prae_binary:fig_PSFimage_method} and in Fig.\ \ref{L0_Prae_binary:fig_PSFimage_binaries}, 
respectively. We sample the parameter space in steps of 0.1 pixel, 3 degrees, and 0.05 for the zero points, 
positions, angle, and flux ratio, respectively. For the best solution, we derive the centroids of every 
source, using the flux values above 50\% of the peak. The positions of the two centroids were employed to 
determine the separation and the position angle of the nodes corresponding to the binary components. 
We also determine the FWHM of the reconstructed PSF\@. In every case we used the best five images, deriving 
the mean and standard deviation values of each output parameter.

\subsection{Physical separation and orientation}
\label{Prae_L0_bin:param_Sep}

We apply this method to four blocks of individual OSIRIS images independently: one block taken on
20 November 2018 and three collected on 24 January 2019\@. 
We measure consistent separations of 6.28, 5.92, 6.08, and 6.00 pixels for the four images, given us the 
confidence that UGCS0845 is a binary with a median projected separation of 6.07$\pm$0.13 pixels (1$\sigma$).

Adopting the plate scale of 9.9407$\pm$0.0071 mas measured in Section \ref{Prae_L0_bin:AO_LGS} for OSIRIS,
we derive a mean physical separation of 60.3$\pm$1.3 mas with a position angle of 295$\pm$1.5 degrees
from North to East anti-clockwise,
and a flux ratio of 0.89$\pm$0.04 (Fig.\ \ref{L0_Prae_binary:fig_PSFimage_binaries}).
This corresponds to a projected physical separation of 11.2$\pm$0.7 au, assuming a distance of
186.18 pc and a dispersion of 10.7 pc for Praesepe (Table \ref{tabPrae_L0_bin:table_param}).
The error bars quoted on angular separations measured in mas take into account the dispersion 
of the astrometric determinations plus the uncertainty of the pixel size determination.

Our method is able to separate the two components of the binary, which we barely resolve.
The diffraction limit of Keck/OSIRIS corresponds to 5 pixels and a FWHM of $\sim$7 pixels 
can be achieved in the best conditions \citep[e.g.][]{liu08b,dupuy12}. 
Our method gives a FWHM of 7.7$\pm$0.8 pixels.
Given the consistency obtained for the separations derived from four independent images
and the similar elongation between the stacked images collected in November 2018 and January 2019,
we are confident that the results are reliable even though AO images often suffer from uncontrolled
systematics.

We apply the same procedure to the OSIRIS images of UGCS0830\@. 
We separate the two components into a binary with a flux ratio of 0.49$\pm$0.07 (1$\sigma$) 
and a projected separation of 6.23$\pm$0.26 pixels (1$\sigma$) equivalent to 61.9$\pm$2.6 mas 
and corresponding to a physical separation of 11.5$\pm$0.8 au (Fig.\ \ref{L0_Prae_binary:fig_PSFimage_binaries}).
We measured the orientation of the binary on the image of 73$\pm$4 degrees, corresponding to a
position angle of 197$\pm$4 degrees.

We repeated the measurements for the NIRC2 images of UGCS0830\@. We selected the
best five images to infer a flux ratio of 0.46$\pm$0.03, separation of 6.27$\pm$0.09 pixels (1$\sigma$)
equivalent to 62.5$\pm$0.9 mas. We measured an angle of 150.4$\pm$1.2 degrees on the image,
translating into an position angle of 180.4$\pm$1.2 degrees.
The quoted error bars stand for the dispersion of the measurements on the selected individual images.
In Table \ref{tabPrae_L0_bin:table_param}, we give the measurements at 1$\sigma$ for both instruments.

We caution, however, that the angular separation may be underestimating
the true projected separation of the binary. Indeed, the two components of UGCS0830 are close and the significantly different
brightness of the two members artificially makes the centroid of the secondary shift towards the position
of the primary. The position angle should not be affected significantly but the flux ratio and separation are.
Nonetheless, we observe that the position angles measured on both instruments for UGCS0830 differ
by 16--17 degrees. 

Despite the excellent agreement in the separation and flux between the two instruments, 
the position angle differs.
At this stage, we fail to reconcile the position angles of UGCS0830 derived from two sets of
observations with two different instruments.
We discuss possible reasons for such difference. (i) The binary might show a significant orbital
motion, which may hint at a very elliptical orbit, or its true separation is smaller than the projected
separation creating a shorter orbital period. However, we do not observe a significant change in the
separation of both components. We conclude that further observations are needed to confirm or not
this possible orbital motion. (ii) The LGS-AO observations suffered from uncontrolled problems
although this hypothesis is not supported by the good quality of the images of the tip-tilt star observed
just before our targets. The AO correction of both datasets is of good quality.
(iii) The determination of the position angles using our method 
(Section \ref{Prae_L0_bin:param_Method}) might suffer from uncontrolled systematics,
but the determination of the angles with our method is robust and confirmed on individual images
as well as with telescope offsets in the headers.
(iv) The companion candidate might be an unrelated field/background source. Considering the
separations of 61.9 mas and 62.5 mas on 11 March 2019 and 22 March 2020 (i.e.\ slightly less than 1 year
apart), the change in position angle from 197 degrees to 180 degrees could be interpreted 
as a relative proper motion of the companion candidate relative to UGCS0830 
of $+$18.3$\pm$5.2 mas/yr and $-$3.4$\pm$3.9 mas/yr in right ascension and declination, respectively.
Hence, the proper motion of the possible companion would be ($-$9.3$\pm$7.3, $-$9.9$\pm$6.3) mas/yr,
2$\sigma$ away from the proper motion of UGCS0830
\citep[($-$27.6$\pm$4.9, $-$6.5$\pm$4.9) mas/yr;][]{boudreault12,lawrence13}.
Furthermore, the proper motion of the companion lies within 3$\sigma$ of the mean
motion of Praesepe \citep[($-$34.2$\pm$2.7, $-$7.4$\pm$4.2) mas/yr;][]{boudreault12},
thus, cannot be totally discarded as a probable member.
We have also computed the probability of finding a source fainter than UGCS0830 ($K$\,=\,16.5 mag) 
and as faint as the companion candidate ($K$\,=\,17.5 mag) within 1 arcsec from UGCS0830\@. 
We estimated the density of objects selecting all point-like and extended sources in the
UKIDSS GCS database in an area of 1 deg$^{2}$ centered around UGCS0830\@. We find 4075
satisfying those criteria, translating into a probability of chance alignment of 10$^{-3}$.
This estimate is an upper limit because the companion candidate lies at about 62 mas,
implying a lower probability of chance alignment of 3.8$\times$10$^{-6}$.
We conclude that the likelihood that both UGCS0830 and the possible 
companion are not physically bound is extremely small.

%
%
\begin{table*}
   \centering
   \caption{Main parameters of the two binary systems in the Praesepe open cluster presented in this work.
   	    1$\sigma$ error bars. 
	    $^{a}$: Parameters from OSIRIS; $^{b}$: Parameters from NIRC2\@.
            $^{c}$: UGCS0845 was also observed on 2458507 and 2458563\@.
	    $^{d}$: system luminosity.
           }
   \begin{tabular}{l c c c c c c c c c}
          \hline
	   Name   & BJD  & SpT & Separation & Separation & PA  &  Flux ratio & $\Delta$\,$K$ & Period & $\log$(L/L$_{\odot}$) \cr
	   ---   & days & --- &  mas &    au      & deg &   ---       &    mag        &  year  & dex   \cr
          \hline
	   UGCS0845$^{a}$ & 2458442$^{c}$ & L1.0$\pm$0.5 & 60.3$\pm$1.3 & 11.2$\pm$0.7 & 295.0$\pm$1.5 & 0.89$\pm$0.04 & 0.13$\pm$0.05 & 96.8$\pm$9.0   & $-$3.24$\pm$0.09$^{d}$ \\
	   UGCS0830$^{a}$ & 2458564       &    ---       & 61.9$\pm$2.6 & 11.5$\pm$0.8 & 197.0$\pm$4.0 & 0.49$\pm$0.07 & 0.77$^{+0.17}_{-0.15}$ & 110.4$\pm$11.9 & $-$3.44$\pm$0.08$^{b}$ \\
	   UGCS0830$^{b}$ & 2458918       &    ---       & 62.5$\pm$0.9 & 11.6$\pm$0.7 & 180.4$\pm$1.2 & 0.46$\pm$0.03 & 0.84$\pm$0.07 & 109.7$\pm$10.5 & $-$3.44$\pm$0.08$^{b}$ \\
          \hline
  \end{tabular}
  \label{tabPrae_L0_bin:table_param}
\end{table*}
%

%
%
\begin{table}
   \centering
   \caption{Photometry of the unresolved systems from Pan-Starrs Data release 1 \citep{chambers16a},
            Sloan \citep{alam15a} in the AB system \citep{fukugita96},  UKIDSS Galactic Clusters 
           Survey (GCS) \citep{lawrence07} in the Mauna Kea Vega system \citep{tokunaga02}, and 
           AllWISE \citep{cutri14} in the Vega system.
           }
   \begin{tabular}{l c c} 
          \hline
          Filter & UGCS0845 & UGCS0830 \cr
          \hline
          PS1 $i$  &  21.536$\pm$0.044 &  21.676$\pm$0.147 \\
          PS1 $z$  &  20.324$\pm$0.054 &  20.682$\pm$0.043 \\
          PS1 $y$  &  19.311$\pm$0.051 &  19.842$\pm$0.060 \\
          SDSS DR12 $r$ &  23.934$\pm$0.443 &  23.823$\pm$0.619 \\
          SDSS DR12 $i$ &  21.878$\pm$0.150 &  21.656$\pm$0.141 \\
          SDSS DR12 $z$ &  19.954$\pm$0.105 &  20.116$\pm$0.145 \\
          UKIDSS GCS $Z$ & 19.783$\pm$0.131 &  19.923$\pm$0.099 \\
          UKIDSS GCS $Y$ & 18.376$\pm$0.055 &  19.076$\pm$0.066 \\
          UKIDSS GCS $J$ & 17.419$\pm$0.033 &  17.868$\pm$0.040 \\ 
          UKIDSS GCS $H$ & 16.689$\pm$0.023 &  17.115$\pm$0.041 \\
          UKIDSS GCS $K$ & 16.033$\pm$0.024 &  16.512$\pm$0.048 \\
          AllWISE $w1$ & 15.647$\pm$0.048 &  16.146$\pm$0.086 \\
          AllWISE $w2$ & 15.318$\pm$0.115 &  15.909$\pm$0.222 \\
          \hline
  \end{tabular}
  \label{tabPrae_L0_bin:photometry}
\end{table}
\subsection{Spectral type}
\label{Prae_L0_bin:param_SpType}
%


UGCS0845 was classified as an L0.4$\pm$0.3 dwarf (equivalent to an effective temperature of 2279$\pm$371\,K) 
comparing its GTC/OSIRIS low-resolution optical spectrum with M and L dwarf templates observed with 
the same instrumentation and inferring spectral indices, see \citet*{boudreault13} for a detailed
description of the method. We revise the spectral classification here, classifying the unresolved system
in the optical and near-infrared independently. We classify the UGCS0845 system as an L1.0$\pm$0.5 dwarf 
(Table \ref{tabPrae_L0_bin:table_param}) by direct
comparison with Sloan optical spectra of old high-gravity L dwarfs \citep{schmidt10b} as shown in the 
left-hand side panel of Fig.\ \ref{L0_Prae_binary:fig_spectra}. We classify the system as an L1.5 dwarf in the 
near-infrared because the VLT/X-shooter spectrum is best fitted by the SpeX spectrum of 2MASS\,J14182962$-$3538060
(right-hand side panel of Fig.\ \ref{L0_Prae_binary:fig_spectra}) classified as a field L1.5 dwarf 
by \citet{kirkpatrick10}\footnote{Spectrum publicly available at http://pono.ucsd.edu/$\sim$adam/}.
This spectral type agree with the one obtained in \citet{manjavacas20a}.
Hence, we conclude that the full spectral energy distribution of UGCS0845 is most consistent with
an L1.0--L1.5$\pm$0.5 dwarf, slightly later than the original classification. According to the
polynomials for old field M6--T9 dwarfs defined by \citet{filippazzo15}, we infer a mean effective temperature
of 2030--2100\,K for the system with an rms of 113\,K\@. Similar effective temperatures (2200$\pm$100\,K) are 
derived from the spectral types-effective temperature relations of \citet{golimowski04a} and \citet{vrba04}.
Because UGCS0845 is an almost equal-flux binary, we would expect the primary and secondary to have
very similar spectral types oround L1.0\@.



%
We did not collect a spectrum for UGCS0830 but its position in the colour-magnitude diagrams
(Fig.\ \ref{L0_Prae_binary:fig_CMDs}) suggests that it is of later spectral type than UGCS0845\@.
Indeed, UGCS0830 appears fainter in $K$-band magnitudes than UGCS0845 and also harbours a redder $Y-K$ colour.
In Section \ref{Prae_L0_bin:param_Lbol}, we infer luminosity intervals for each component of 
UGCS0830\@. Using the luminosity-spectral type relation of \citet{filippazzo15} valid for old field M6--T9 
dwarfs, we estimate spectral types of M9--L1 and L3--L5 for the primary and secondary of UGCS0830, respectively
(Table \ref{tabPrae_L0_bin:table_param}).

\subsection{Bolometric luminosity}
\label{Prae_L0_bin:param_Lbol}
%


We integrate the spectral energy distribution of the UGCS0845 unresolved system using photometry from the
three distinct public surveys (Table \ref{tabPrae_L0_bin:photometry}): the $riz$ magnitudes from the
Sloan Digital Sky Survey DR12 \citep{alam15a}, the $ZYJHK$ magnitudes from the UKIDSS Galactic Clusters
Survey \citep{lawrence07}, and the AllWISE survey \citep{wright10,cutri14}. We complement the photometry
with the spectrum of the BT-Settl model at a temperature of 2100\,K (with solar-metallicity and
gravity of $\log$(g)\,=\,5.0 dex) to estimate fluxes in the blue (negligible for these objects below 300 nm)
and the red part of the spectral energy distribution (about 50\% beyond 5 microns). We derive a total
(i.e.\ for the unresolved system) photometric bolometric luminosity of 5.616$\times$10$^{4}$ erg/s
for UGCS0845, yielding $\log$(L/L$_{\odot}$)\,=\,$-$3.25$\pm$0.03 dex, assuming a distance of 186.18 pc.
The depth of the cluster taken as the tidal radius adds a non-negligible error budget of 0.06 dex,
yielding $\log$(L/L$_{\odot}$)\,=\,$-$3.25$\pm$0.07 dex (Table \ref{tabPrae_L0_bin:table_param}).

We also estimate the bolometric luminosity of UGCS0845 by integrating the full flux-calibrated optical and
near infrared spectrum, and the $W1$ and $W2$ photometry in the mid-infrared. We calibrate in flux the
optical and the near-infrared spectra independently, using the $z$-Pan-Starrs and the $J$-UKIDSS photometry,
respectively. To obtain the bolometric luminosity, we integrate the full spectral energy distribution
from 0 to 1000 microns, interpolating the gaps between the available optical and near-infrared spectra,
and photometric points with their associated errors by performing a linear interpolation
(see Manjavacas et al.\ 2020).
We find $\log$(L/L$_{\odot}$)\,=\,$-$3.24$\pm$0.09 dex, in agreement with the photometric value.
Given the flux ratio of 0.89 with an uncertainty of 0.04, which translates into an additional budget
uncertainty of 0.02 dex, we infer luminosities of $\log$(L/L$_{\odot}$)\,=\,$-$3.49$\pm$0.09 dex
and $-$3.60$\pm$0.09 dex for the primary and secondary, respectively (Table \ref{tabPrae_L0_bin:table_param}),
adopting equal spectral energy distribution as reasonable approximation for UGCS0845\@.
A more accurate distance for the binary system would significantly decrease the uncertainty on
the luminosities and, more importantly, on the masses but $Gaia$ data are not available for our sources.


We have only photometry for UGCS0830 (Table \ref{tabPrae_L0_bin:photometry})
so we integrate the spectral energy distribution of the unresolved system using photometry from
Sloan, UKIDSS GCS, and AllWISE, as for UGCS0845\@. We infer a system bolometric luminosity of 
$\log$(L/L$_{\odot}$)\,=\,$-$3.44$\pm$0.08 dex. Given our adopted flux ratio of 0.46$\pm$0.03, 
we infer luminosities of $\log$(L/L$_{\odot}$)\,=\,$-$3.60$\pm$0.08 dex
and $-$3.94$\pm$0.10 dex for the primary and secondary, respectively.
We note that we did not take into account the difference in $K$-band bolometric correction 
between the primary and secondary components of UGCS0830 that can reach up to 0.2 mag 
between a field M9 and L5 dwarf \citep{filippazzo15}.

%
%
%
\begin{table*}
 \centering
 \caption{
Masses of each component of the UGCS0845 (top two rows) system as a function of the flux ratio 
(0.89$\pm$0.04) and the age assumed for Praesepe (600, 700, and 800 Myr) with an upper limit of 1 Gyr. 
The formal uncertainty on each mass estimate is of the order of 0.001 M$_{\odot}$.
We list the masses of each component of UGCS0830 for the adopted flux
ratio and the age range of the cluster  in the bottom two rows.
}
 \begin{tabular}{@{\hspace{0mm}}c|ccc|ccc|ccc|ccc@{\hspace{0mm}}} 
  \hline
  \hline
Age (Myr)                    & \multicolumn{3}{|c|}{600} & \multicolumn{3}{c}{700} & \multicolumn{3}{c}{800} & \multicolumn{3}{c}{1000}  \\
  \hline
Flux ratio                   & 0.85 & 0.89 & 0.93 & 0.85 & 0.89 & 0.93 & 0.85 & 0.89 & 0.93 & 0.85 & 0.89 & 0.93 \\
  \hline
M$_{\rm prim}$ (M$_{\odot}$) &  0.074 & 0.074 & 0.074 & 0.076 & 0.076 & 0.076 & 0.078 & 0.078 & 0.078 & 0.080 & 0.080 & 0.080 \\
M$_{\rm sec}$ (M$_{\odot}$)  &  0.072 & 0.072 & 0.072 & 0.074 & 0.074 & 0.074 & 0.076 & 0.076 & 0.076 & 0.078 & 0.078 & 0.078 \\
 \hline
 \hline
Age (Myr)                    & \multicolumn{3}{|c|}{600} & \multicolumn{3}{c}{700} & \multicolumn{3}{c}{800} & \multicolumn{3}{c}{1000}  \\
 \hline
Flux ratio                   & 0.43 & 0.46 & 0.49 & 0.43 & 0.46 & 0.49 & 0.43 & 0.46 & 0.49 & 0.43 & 0.46 & 0.49 \\
 \hline
M$_{\rm prim}$ (M$_{\odot}$) &  0.071 & 0.071 & 0.070 & 0.074 & 0.073 & 0.073 & 0.076 & 0.075 & 0.078 & 0.078 & 0.077 & 0.077 \\
M$_{\rm sec}$ (M$_{\odot}$)  &  0.051 & 0.055 & 0.056 & 0.058 & 0.059 & 0.060 & 0.061 & 0.063 & 0.065 & 0.066 & 0.067 & 0.067 \\
 \hline
 \hline
 \end{tabular}
 \label{tabPrae_L0_bin:table_masses}
\end{table*}
\subsection{Masses}
\label{Prae_L0_bin:param_Mass}
\subsubsection{UGCS0845}
\label{Prae_L0_bin:param_Mass_UGCS0845}

We derive the masses from the total bolometric luminosity calculated from the full spectral energy
distribution using the latest BT-Settl isochrones \citep{baraffe15}. For a flux ratio of 0.76 and a
mean age of 700 Myr, we infer masses of 0.078 M$_{\odot}$ and 0.073 M$_{\odot}$ for the primary and 
secondary, respectively (Table \ref{tabPrae_L0_bin:table_masses}). The uncertainty of 0.07 dex in 
$\log$(L/L$_{\odot}$), which does not include the uncertainty in the differing spectral energy distributions, 
translates approximately into a (formal) uncertainty of at most 0.002 M$_{\odot}$ on the mass.
We show the full range of values taking into account all uncertainties on flux ratio and age in
Table \ref{tabPrae_L0_bin:table_masses}.

Assuming ages of 600 Myr and 800 Myr for Praesepe, the masses would change by 0.002 M$_{\odot}$.
In the case of flux ratios of 0.85 and 0.93,
the formal differences in the masses of the primary and the secondary is below one Jupiter mass 
(Table \ref{tabPrae_L0_bin:table_masses}). We have also included the masses for an upper limit of 
1 Gyr on the age of Praesepe.

In all cases, the secondary straddles the stellar/substellar boundary (0.072--0.076 M$_{\odot}$)
and might be either a very low-mass star or a brown dwarf depending on the age adopted for the cluster,
while the primary would be a very low-mass star (0.074--0.078 M$_{\odot}$) just above the hydrogen-burning 
limit set to 0.072 M$_{\odot}$ at solar metallicity \citep{chabrier97}.

To obtain an independent estimate of the mass of the system with the lithium test in a similar-aged 
cluster such as the Hyades, we have compared the magnitudes of UGCS0845 with the very low mass stars 
and brown dwarfs confirmed spectroscopically in the Hyades open cluster 
\citep[d\,=\,47.50$\pm$0.15 pc;][]{babusiaux18} whose age is comparable to Praesepe
\citep[600--750 Myr;][]{maeder81,mermilliod81,mazzei88,deGennaro09,lebreton01,brandt15a,martin18a,lodieu18b}.
Among the 12 L dwarfs identified in the Hyades by \citet{hogan08}, ten of them were confirmed
spectroscopically as members \citep{lodieu14b}, including three brown dwarfs with masses below
0.06 M$_{\odot}$ with lithium in absorption at 6707.\,8\,\AA{} \citep{martin18a}.
The lithium depletion boundary in the Hyades is located at M$_{J}$\,=\,12.2--12.7 mag and 
M$_{K}$\,=\,10.8--10.9 mag, respectively. The absolute magnitudes of UGCS0845 are
M$_{J}$\,=\,11.07$\pm$0.10 and M$_{K}$\,=\,9.68$\pm$0.11 mag, where the uncertainty take into account
the depth of the cluster (Table \ref{tabPrae_L0_bin:photometry}).


In the case of a system with the secondary having 0.89 times the flux of the primary, the magnitudes 
of the primary and secondary would be M$_{K}$\,=\,10.38$\pm$0.05 mag and M$_{K}$\,=\,10.50$\pm$0.09 mag,
respectively.

According to the latest BT-Settl models \citep{baraffe15}, the hydrogen-burning limit is 1.0 mag and 0.6 mag 
brighter than the lithium depletion boundary in the $J$- and $K$-band, respectively, at an age of 600 Myr 
(1.1 and 0.8 mag at 700 Myr), suggesting that the primary might be a star and the secondary a high-mass brown dwarf.
Hence, we do not expect a strong lithium abundance for high-mass brown dwarfs because they will 
most likely have depleted most of the original lithium at the age of the cluster. However, 
if we can detect a tiny fraction 
of lithium in the spectrum of the secondary, it will be very important to constrain the masses, 
evolutionary models and even Li destruction models when dynamical masses are in hands.

Applying the Kepler's third law, we infer an orbital period of 96.8$\pm$9.0 years, 
assuming total mass of 0.150$\pm$0.004 M$_{\odot}$ and zero eccentricity (Table \ref{tabPrae_L0_bin:table_param}).

\subsubsection{UGCS0830}
\label{Prae_L0_bin:param_Mass_UGCS0830}

We derive the masses from the total bolometric luminosity calculated from the photometry using the 
BT-Settl isochrones. For a flux ratio of 0.46 and a
mean age of 700 Myr, we infer masses of 0.073 M$_{\odot}$ and 0.059 M$_{\odot}$ for the primary and 
secondary, respectively (Table \ref{tabPrae_L0_bin:table_masses}). The uncertainty of 0.03 dex in 
$\log$(L/L$_{\odot}$) translates approximately into a (formal) uncertainty of at most 
0.002 M$_{\odot}$ on the mass.
Repeating the procedure for the 600$-$800 Myr range of possible ages for Praesepe 
(Table \ref{tabPrae_L0_bin:table_masses}), 
we infer masses close to or above the hydrogen-burning limit for the primary (0.070--0.078 M$_{\odot}$) and 
in the substellar regime for the secondary (0.051--0.065 M$_{\odot}$), 
implying that it would be the first brown dwarf discovered in the Praesepe with L spectral type 
for which a precise dynamical mass can be derived in the next decades. 
If Praesepe is 1\,Gyr-old, the masses of the primary and secondary of UGCS0830 could be as high
as 0.078 and 0.066 M$_{\odot}$, respectively.

We also performed the computation taking the Hyades as a reference, as for UGCS0845\@.
We inferred an absolute $K$-band magnitude of M$_{K}$\,=\,10.16$\pm$0.08 mag for UGCS0830\@.
If we split the system into a binary with a flux ratio of 0.46, we infer absolute magnitudes of 
M$_{K}$\,=\,10.573$^{+0.199}_{-0.192}$ mag and 11.415$^{+0.295}_{-0.245}$ mag for the primary and secondary, respectively,
taking into account the depth of the cluster and propagating uncertainty on the photometry and flux ratio.
Hence, the primary is expected to be a massive brown dwarf whose
lithium has been depleted while the secondary is a brown dwarf at the lithium depletion 
boundary. For a brown dwarf that has fully retained its lithium, we expect a pseudo-equivalent
widths in the 6--20\,\AA{} range \citep{kirkpatrick00,kirkpatrick08,cruz09,martin18a,lodieu18b}.
Taking into account the dilution factor due to the fact that the primary should not exhibit lithium
in absorption, we would expect a system pseudo-equivalent width of 1.0--7.6\,\AA{} for the lithium
absorption line at 6707.8\,\AA{} in the integrated optical spectrum of the system.

Applying the Kepler's law, we infer an orbital period of 109.7$\pm$12.4 years, 
assuming zero eccentricity and a total mass of 0.132$\pm$0.008 M$_{\odot}$ 
(Table \ref{tabPrae_L0_bin:table_param}).

%
%
\begin{figure}
  \centering
	\includegraphics[width=\linewidth, angle=0]{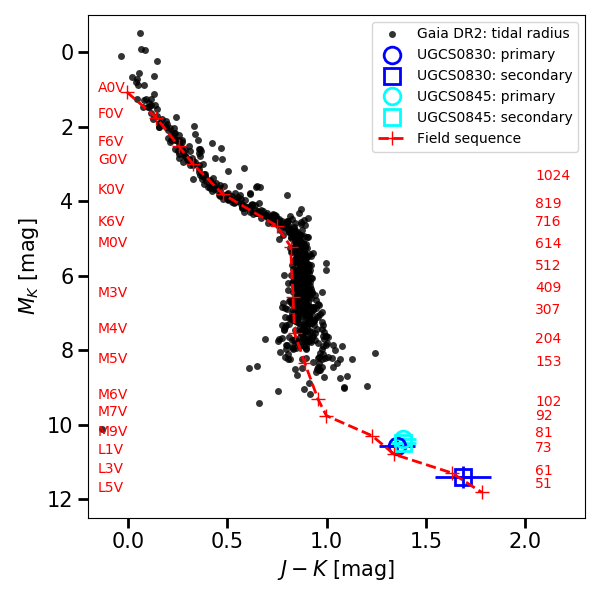}
   \caption{
Hertzsprung-Russell diagram for Praesepe. We plot cluster members from $Gaia$ DR2
within the tidal radius \citep{lodieu19b} as black dots. We overplot the position of 
each component of UGCS0845 and UGCS0830 in cyan and blue, respectively.
We added an estimate of the spectral type from the absolute magnitude-spectral relation
of field dwarfs \citep{pecaut13} and masses predicted by the BT-Settl models for
an age of 700 Myr \citep{baraffe15}.
   }
   \label{L0_Prae_binary:fig_HRdiag}
\end{figure}
%

%
%
\section{Discussion}
\label{Prae_L0_bin:discussion}

We have confirmed the binarity of UGCS0845 and a companion candidate to UGCS0830.
In this discussion, we assume that both are true binaries. Both systems represent the first binaries across 
the hydrogen-burning limit in Praesepe and the first L--L binaries in this cluster. Many surveys 
referenced throughout this work have screened low-mass stars and brown dwarfs for companions
in clusters younger than 700 Myr, but only L--L binaries have been reported in the Hyades so far.
These new systems represent important discoveries to measure dynamical masses across the hydrogen-burning
limit and in the substellar regime, locate the lithium depletion boundary, provide an estimation
of the multiplicity properties of low-mass stars and brown dwarfs at a given age, and constrain
stellar evolutionary models in a mass regime where physics is complex \citep{baraffe15}.

We plot the H-R diagram of Praesepe in Fig.\ \ref{L0_Prae_binary:fig_HRdiag} showing all members
within the tidal radius of the cluster identified in $Gaia$ DR2 \citep{lodieu19b} 
with the location of the components of the two binaries presented
in this work. We observe that they lie across the hydrogen-burning limit based on model predictions
\citep{baraffe15} and have spectral types later than M9/L0 according to the field relation of
\citet{pecaut13}. We have only the $K$-band magnitude difference for both systems so we assumed
the $J-K$ colour of the system for each component of UGCS0845 because the almost equal flux ratio.
In the case of UGCS0830, we assume the colour of a L3 dwarf for the secondary and included an
additional error of $+$0.1 mag to take into account the lack of spectral type.
These two binaries represent important to extend the Praesepe sequence into
the substellar regime with future deep surveys.

\citet{boudreault12} estimated a photometric multiplicity of 23.2$\pm$5.6\% in the 0.4--0.1 M$_{\odot}$ 
mass range. This is comparable to the frequency of Hyades low-mass stars and brown dwarfs derived
early on by \citet*{gizis95a} with a fraction of 27$\pm$16\% and later
by \citet{duchene13a} who resolved 3 out of 16 targets with adaptive optics (19$^{+13}_{-6}$\% in 2--350 au).
However, this estimate should be revised because one of the resolved system (Hya05) was rejected as a 
spectroscopic member of the Hyades \citep{lodieu14b}, yielding a lower multiplicity of $\sim$13.3\% (2/15).
The binary sequence of Praesepe is well seen in various colour-magnitude diagrams and fitted by a
red line in Fig.\ \ref{L0_Prae_binary:fig_CMDs}. We confirm two of the photometric binary candidates 
with adaptive optics, suggesting that the multiplicity of 23.2$\pm$5.6\% estimated by \citet{boudreault12}
for the 0.1--0.07 M$_{\odot}$ range may hold, if all photometric and astrometric candidates are later 
confirmed as spectroscopic members. This photometric binary fraction agrees with the overall multiplicity
fraction of field brown dwarfs \citep*[18--28\%][]{duchene13b} and is also consistent with the
theoretical predictions of hydrodynamical simulations \citep[27.3$\pm$11.6\%;][]{bate12}.

For comparable mass intervals but younger ages, where multiplicity fractions remain very uncertain, 
we should highlight some key results. Considering current uncertainties sample sizes, those
comparisons should be taken with a pinch of salt but necessary to understand the evolution of
multiple systems with age. In the Pleiades, whose age is around 125 Myr \citep{stauffer98,barrado04b}, 
\citet{martin03} and \citet{bouy06a} derived a binary fraction of 9--27\%
for separations larger than 7 au and mass ratios greater than 0.45--0.9 based on a sample of
15 substellar members with spectral types in the M6--M9 range (0.055--0.065 M$_{\odot}$).
At younger ages (i.e.\ star-forming regions younger than about 10 Myr), there might be a trend
for a sharp decrease from 15--28\% for very low-mass stars (0.15--0.07 M$_{\odot}$) down to
less than 11\% (50\% confidence) below 0.1 M$_{\odot}$ \citep{kraus12a,duchene13b}. 
The latter is consistent with the 3--7\% of binaries imaged in a sample of $>$100 young M6 dwarfs 
and later type in Taurus and Chamaeleon \citep{todorov14}.
The binary fraction of brown dwarfs in Praesepe might not be primordial but affected by dynamical
evolution of the cluster.



%
%
\section{Conclusions and future plans}
\label{Prae_L0_bin:conclusions}

We report two new ultracool binary systems composed of two L dwarfs straddling the stellar/substellar 
in the Praesepe open cluster.
Both systems have projected separations of about 60 mas, corresponding to physical separations
of $\sim$11--12 au and minimum orbital periods of approximately 100 years. If confirmed as
a companion, the secondary of the
system with the largest flux ratio lies unambiguously in the brown dwarf regime, and we argue
that the lithium feature should be present in absorption at 6707.8\,\AA{}.
Both systems are key to locate with higher precision the position of the
stellar/substellar and lithium depletion boundaries in Praesepe.
If the photometric multiplicity among Praesepe members with masses below the hydrogen-burning
limit holds, we argue that the binary fraction in Praesepe might not be primordial.

We plan to continue the follow-up of these two binaries with the LGS systems on Keck and the future 
AO laser guide star system on the Gran Telescopio de Canarias \citep[GTCAO-LGS;][]{bejar19}
and the FRIDA instrument \citep[inFRared Imager and Dissector for Adaptive optics;][]{watson16a} 
for the next decade to derive dynamical masses of both components and measure the amount 
of lithium in each component. The accuracy of these dynamical
masses will depend on the evolution of the separation of the system with time, i.e.\ whether
we observed the system at periastron or not. The upcoming HARMONI spectrograph 
\citep[High Angular Resolution Monolithic Optical \& Near-infrared Integral field;][]{tecza09a,thatte10a}
on the extremely large telescope should provide us with better spatial resolution and sensitivity to 
monitor their orbits. 

%
%
\section*{Acknowledgements}
NL and VJSB were supported by the Spanish Ministry of Economy and
Competitiveness (MINECO) under the grants AYA2015-69350-C3-2-P\@.
CdB acknowledges the funding of his sabbatical position through the
Mexican national council for science and technology (CONACYT grant CVU
No.\ 448248). He is also thankful for the support from the Jesus Serra Foundation Guest Program.
MRZO was supported by MINECO grant number AYA2014-54348-C3-2-R\@.
The data presented herein were obtained at the W. M. Keck Observatory, which is operated as a scientific 
partnership among the California Institute of Technology, the University of California and the National 
Aeronautics and Space Administration. The Observatory was made possible by the generous financial support 
of the W. M. Keck Foundation. The authors wish to recognise and acknowledge the very significant cultural 
role and reverence that the summit of Maunakea has always had within the indigenous Hawaiian community. 
We are most fortunate to have the opportunity to conduct observations from this mountain. \\
Based on observations made with ESO Telescopes at the La Silla Paranal Observatory under programme 
ID 098.C-0277(A), PI Manjavacas.
This work is based on observations (programme GTC66-12B; PI Boudreault) made with the Gran 
Telescopio Canarias (GTC), operated on the island of La Palma in the Spanish Observatorio 
del Roque de los Muchachos of the Instituto de Astrof\'isica de Canarias. 
This research has made use of the Simbad and Vizier \citep{ochsenbein00}
databases, operated at the Centre de Donn\'ees Astronomiques de Strasbourg
(CDS), and of NASA's Astrophysics Data System Bibliographic Services (ADS). \\
This work has made use of data from the European Space Agency (ESA) mission {\it Gaia} 
(\url{https://www.cosmos.esa.int/gaia}), processed by the {\it Gaia} Data Processing and 
Analysis Consortium (DPAC, \url{https://www.cosmos.esa.int/web/gaia/dpac/consortium}). 
Funding for the DPAC has been provided by national institutions, in particular the institutions
participating in the {\it Gaia} Multilateral Agreement.
    Funding for the Sloan Digital Sky Survey IV has been provided by
    the Alfred P. Sloan Foundation, the U.S. Department of Energy Office of
    Science, and the Participating Institutions. SDSS-IV acknowledges
    support and resources from the Center for High-Performance Computing at
    the University of Utah. The SDSS web site is www.sdss.org.
    SDSS-IV is managed by the Astrophysical Research Consortium for the
    Participating Institutions of the SDSS Collaboration including the
    Brazilian Participation Group, the Carnegie Institution for Science,
    Carnegie Mellon University, the Chilean Participation Group, the French Participation Group,
Harvard-Smithsonian Center for Astrophysics,
    Instituto de Astrof\'isica de Canarias, The Johns Hopkins University,
    Kavli Institute for the Physics and Mathematics of the Universe (IPMU) /
    University of Tokyo, Lawrence Berkeley National Laboratory,
    Leibniz Institut f\"ur Astrophysik Potsdam (AIP),
    Max-Planck-Institut f\"ur Astronomie (MPIA Heidelberg),
    Max-Planck-Institut f\"ur Astrophysik (MPA Garching),
    Max-Planck-Institut f\"ur Extraterrestrische Physik (MPE),
    National Astronomical Observatory of China, New Mexico State University,
    New York University, University of Notre Dame,
    Observat\'ario Nacional / MCTI, The Ohio State University,
    Pennsylvania State University, Shanghai Astronomical Observatory,
    United Kingdom Participation Group,
    Universidad Nacional Aut\'onoma de M\'exico, University of Arizona,
    University of Colorado Boulder, University of Oxford, University of Portsmouth,
    University of Utah, University of Virginia, University of Washington, University of Wisconsin,
    Vanderbilt University, and Yale University. \\
Based on data from the UKIRT Infrared Deep Sky Survey (UKIDSS). The UKIDSS project
is defined in \citet{lawrence07} and uses the UKIRT Wide Field Camera \citep[WFCAM;][]{casali07}.
The photometric system is described in \citet{hewett06} and the calibration is described
\citet{hodgkin09}. The pipeline processing and science archive are described in
Irwin et al.\ (in prep) and \citet{hambly08}. \\
This publication makes use of data products from the Wide-field Infrared Survey Explorer, which
is a joint project of the University of California, Los Angeles, and the Jet Propulsion Laboratory/California
Institute of Technology, and NEOWISE, which is a project of the Jet Propulsion Laboratory/California
Institute of Technology. WISE and NEOWISE are funded by the National Aeronautics and Space Administration.

%
%

{\bf{Data Availability Statement}}
The data underlying this article will be shared on reasonable request to the corresponding author.

%
%
\bibliographystyle{mnras}
\bibliography{../../AA/mnemonic,../../AA/biblio_old}



%
%
%


\bsp	
\end{document}